\documentclass{ws-p8-50x6-00}
\usepackage{epsfig}
\usepackage{psfrag}
\newcommand{\nc}{\newcommand}
\nc{\cL}{{\cal L}}
\nc{\cO}{{\cal O}}
\nc{\tr}{{{\rm tr}\,}}

\nc{\bk}{{{\bf k}}}
\nc{\bx}{{{\bf x}}}

\nc{\simo}[1]{{\stackrel{#1}{\simeq}}}
\nc{\geqo}[1]{{\stackrel{#1}{\geq}}}
\nc{\geo}[1]{{\stackrel{#1}{>}}}
\nc{\guo}[1]{{\stackrel{#1}{\succ}}}

\nc{\rbo}{\raisebox}
\nc{\RR} {\rangle \! \rangle}
\nc{\LL} {\langle \! \langle}
\nc{\rmi}[1]{{\mbox{\small #1}}}
\nc{\eq}{eq.~}
\nc{\nr}[1]{(\ref{#1})}
\nc{\ul}{\bf }
\nc{\mc}{\multicolumn}
\nc{\todo}[1]{\par\noindent{\bf $\rightarrow$ #1}}

\nc{\cu}{{\cal u}}

\begin{document}

\title{High Temperature QCD and Dimensional Reduction}

\author{Bengt Petersson}

\address{University of Bielefeld, Faculty for Physics, P.O.Box 10 01 31,
D-33501 Bielefeld, Germany\\
E-mail: bengt@physik.uni-bielefeld.de}


\maketitle

\section{Abstract}

In this talk I will first give a short discussion of some lattice
results for QCD at finite temperature. I will then describe
in some detail the technique of dimensional reduction, which in
principle is a powerful technique to obtain results on the long
distance properties of the quark-gluon plasma. Finally I will
describe some new results, which test the technique in a simpler
model, namely three dimensional gauge theory.

\section{Introduction}

The physical challenges for lattice computations of high temperature
QCD are to obtain information, which can be of use for the planning and
analysis of recent experiments, as well as to give theoretical insight
into the non perturbative properties of QCD. Some of the more important
goals are the determination of the value of the transition temperature
in physical units, the equation of state, the order or even the
existence of a phase transition to a high temperature quark-gluon
plasma, and the long distance properties of the plasma phase.

A simple picture of the finite temperature properties of QCD, as first
discussed by Cabibbo and Parisi \cite{cabibbo}, predicts a low temperature 
confined phase, and a high temperature phase which is essentially an ideal
quark-gluon gas. The transition temperature is expected to be of the
order of the Hagedorn temperature, i.e. around $140$ MeV.
Up to now only lattice calculations can give
more quantitative information. Reviews of the lattice results can be found in
the proceedings of the yearly lattice conferences, the most recent published
review is reference \cite{karsch}. Here I will only emphasize some of the 
results on the quantities mentioned above.

In the pure gluon theory it has been shown that the thermodynamics can
be essentially completely solved, by continuum extrapolation of the
lattice results.
In this way, one can estimate the ratio between the critical temperature
and the square root of the string tension to be $0.630 (5)$.\cite{boyd}
 A similar extrapolation using a different lattice
 action gives  $0.650 (5)$.\cite{okamoto} The difference seems to come from the
measurement of the string tension, which is difficult to estimate precisely,
and where systematic errors come from the uncertainties in the non asymtotic
terms in the spatial distance. Still, the estimates agree within 3 \%.
Using a value $\sqrt{\sigma} = 425$ MeV, one obtains a
rather high critical temperature of around 270 MeV. 

Unfortunately, as far as the order of the transition and the critical 
temperature
goes, the pure gluon theory is not a good approximation to full QCD with
dynamical quarks. Direct lattice calculations with 
 systematic continuum extrapolation in this case are
far too time consuming for present computers, and would, with
the methods currently available, need computing power in the range of
10 to 100 Teraflops.
From the present results one expects the critical temperature
e.g. in ratio to the string tension to be considerably lower. In physical
units one estimates the critical temperature to be  in the range of 
150 - 190 MeV,
where the latest results are near 170 MeV.\cite{karsch,ali}
 It is difficult to estimate the systematic errors.
The order of the transition may be very sensitive to the strange quark. A
crossover as a shadow of a nearby second or first order transition is
favoured by the data, and also by general universality arguments.

 In the pure gluon theory, the pressure and energy density as a function of
temperature can also be extrapolated to the continuum limit, 
and also here results from different calculations agree
within about 3\%.\cite{boyd,okamoto}
 The main message from these calculations is that
the free gluon gas limit is approached rather slowly, and that even at
temperatures of two times the critical temperature, which corresponds
to energy densities at least 16 times the energy density at the transition,
there are 15 - 20\% corrections to the ideal gas limit. Qualitatively, one
observes a similar behaviour in full QCD.\cite{laermann} 
In this case a continuum 
extrapolation
has not been attempted, and one is restricted to results on rather coarse
lattices. One should remark that the experiments planned at Brookhaven and even
LHC will operate in a range of temperatures, where the deviations from the
ideal gas behaviour should be substantial.
 A better understanding
of the non perturbative effects is certainly needed. Resummation techniques
of the perturbative series, essentially introducing a gluon mass, do agree
with the lattice results for sufficiently high temperature 
(from about $2 T_c$).
\cite{blaizot,andersen}
The systematic errors in this procedure are also not very well
known, and futher investigation of the long distance properties of 
the plasma is certainly needed. Dimensional reduction is a technique, which 
is well suited for studying these long range properties. In the following 
I will
give a short review of the general technique, and report on some recent 
results, which test the approach in detail.\cite{us}

\section{Dimensional Reduction}
Dimensional reduction is in principle a powerful technique that
was introduced in the early eighties as a mean to treat
high temperature field theories. 
In the Euclidean
formulation such theories are defined in a volume with one
compact dimension of extent $1/T$. As the temperature T becomes
large, one may expect that the non-static modes in the
temperature direction can be neglected, and that one is left
with a theory in one dimension less \cite{ginsparg,appelquist}.
This naive reduction is, however, only exact in the classical
limit. In general, one has to integrate over the nonstatic
modes to obtain the effective action \cite{nadkarni,landsman,thomas}.
As was shown in Refs
\cite{nadkarni,thomas}
the effective action for the long distance phenomena will, however,
only contain a limited number of local terms at sufficiently high
 temperature, because the integral over the
nonstatic modes does not contain infrared divergencies.
Furthermore, the coefficients can be
determined from a perturbative expansion of the integral
over the non-static modes. This effective action is then expected
to describe correctly the infrared behaviour of the full
theory. It has been successfully applied to the electroweak phase
transition.\cite{kajantie} 
In QCD the action is of the form
\be
S = \frac{1}{g^2} \, \int^{1/T}_0 \, d\tau \, \int \,
d^d x \left[ F_{\mu\nu}^2 (\tau, \vec x ) + \bar\psi D \psi \right]
\ee
where possible ghost and gauge fixing terms have been suppressed.
It is convenient to analyse the theory in the static Landau gauge, defined
by 
\bea
A_0(\tau,\vec{x})=A_0(\vec{x})\\
\int^{1/T}_0  d\tau \, \partial_i A_i = 0 .
\eea
The bosonic and fermionic  fields are periodic and antiperiodic respectively
in the temperature direction, and the corresponding Matsubara frequencies are
\bea
\omega_n & = & 2 \pi n T \,\,\,\,\, n = 0, \pm 1, \pm 2 , ...\,\,\,\,\,\,\,
{\rm Bosons}  \nonumber \\
\omega_n & = & ( 2 n + 1) \pi T \,\,\,\,\,\,\,\,\,\,\,\,\,\,\,\,\,\,\,\,\,\,
\,\,\,\,\,\,\,\,\,\,\,\,\,\,\,\,\,\,{\rm Fermions}
\eea
Note that all modes become infinitely heavy in the high temperature limit,
apart from the static modes in the bosonic case. If one neglects the non
static modes, or at least integrate them out perturbatively, a much simpler
theory describes the long distance properties of the plasma.
At the tree level, one is left with the effective action in one less dimension
\be
S \rightarrow \frac{1}{g^2 T} \, \int \, d^d x \left[ F^{St~2}_{ij}
(\vec x) +
\left( D_i A^{St}_0 (\vec x) \right)^2 \right].
\ee
In the quantum theory, one should integrate over the non static modes. This
gives in principle a very complicated non local effective action. If one
observes that the integration over these modes is infrared finite, one may, however,
develop the action in a series of local terms, which describe the leading
behaviour of correlation functions at large T and small momenta.

If we introduce a field $\phi$ instead of $A_0$
\be
A^{St}_0 (\vec x) = \sqrt T \, \phi (\vec x)
\ee
we can see essentially from power counting and symmetry arguments 
that the dominating terms
in perturbation theory at high temperature and long distances
are the first orders in a polynomial in $\phi$:

\bea
3 + 1 \rightarrow 3 &~~~~~~~~~~   2 + 1 \rightarrow 2 \nonumber \\
 &  \nonumber  \\
g^2 T^2 \,\, \phi^2~~ &~~~~~~~  g^2_3 T \,\,\phi^2  \nonumber \\
 &  \nonumber \\
g^4 T \,\, \phi^4~~~~   &~~~~~~  g^4_3 \,\, \phi^4   \nonumber \\
 & \nonumber  \\
g^6 \,\, \phi^6~~~~~~     &~~~~~~~~  g^6_3 / T \,\, \phi^6  \nonumber 
\eea
The numerical value of the coefficients can be calculated in perturbation 
theory.

Although this is a very nice result, which means essentially that e.g.
full QCD in 3+1 dimensions is replaced by a purely bosonic theory
in three dimensions, the range of validity of the approximation
is not yet known, although several results indicate that it may be
valid down to a temperature of about two times the critical temperature.
\cite{su2,su3,karschpetr,datta,hart}

In a recent work we have investigated how dimensional reduction
works, both for correlations between Polyakov loops, which are
related to chromoelectric screening and for spatial Wilson loops,
which are related to the chromomagnetic sector. We have performed
this investigation on a simpler model, three
dimensional $SU(3)$
gauge theory. Thereby we can obtain a high statistical accuracy
in the comparison. The reduced model is a two dimensional 
adjoint Higgs model. This model is interesting in its own right,
unfortunately it has not been analytically solved.

Three dimensional gauge theories have several similarities with the
full four dimensional case, as e.g. confinement and asymptotic freedom.
They are, however, superrenormalizable, and have infrared singularities,
which are stronger than in four dimensions. Even the leading perturbative
definition of the Debye screening mass is not well defined, because of a
infrared logarithmic singularity.

I will not go into the technical details of the reduction, but instead I 
refer to our publication \cite{us}. 
We have simulated the three dimensional theory
in a wide range of temperature, for lattices with four steps in the temperature
direction, using the 
Wilson action.\cite{lego} We also calculated in the one loop
approximation, the effective action in the lattice regularization.
Again the action can be written in terms of lattice variables. In the
scaling limit we get the continuum form
\bea
 \cL_{eff}&=&\frac {1}{4} \sum _{c=1}^8 F_{ij}^c\,F_{ij}^c + \tr
[D_i\phi]^2 +\frac {g_2^2}{32\pi}\biggl (\frac {g_2}{T}\biggr )^2\,
\tr \,\phi ^4 +\cL_{CT}, \\
 D_i \phi &=& \partial_i \phi + ig_2 [A_i,\phi],
\nonumber \\
 F_{ij} &=& \partial_i A_j - \partial_j A_i + i g_2 [A_i, A_j], \nonumber \\
 \cL _{CT} &=& -\frac {3g_2^2}{2\pi}
 \biggl [-\log (aT)+5/2\log 2-1\biggr ]\,\tr \,\phi ^2. 
\label{counter}
 \eea
This is a 2D, SU(3) gauge invariant Lagrangian for an adjoint scalar
$\phi$, but it is far from being the most general one. The gauge coupling
$g_2 = \sqrt{g_3^2 T}$, with its canonical dimension one in energy, sets
the scale. The non kinetic quadratic term is
the counterterm $\cL_{CT}$, suited to a lattice UV regularization with spacing
$a$. As is explained in our paper \cite{us}, it is very important to include
properly the regularized divergence in the $\phi^2$ term.

On the lattice, the threedimensional model is defined on a $L_0\times L_s^2$ 
lattice
and the two dimensional model on a $L_s^2$ lattice. In the reduced model,
the spatial vector fields $A_i$ are replaced by the group elements $U_i$, as
usual, and the space part of the kinetic term is replaced by the Wilson action.
The field $\phi$ is kept as a scalar field defined on the sites of the lattice,
and the appropriate covariant definition of the covariant derivative is used.
Thereby the explicit $Z_3$ symmetry of the original three dimensional theory is
lost. It can be restored formally, e.g. by substituting $\phi$ by a group 
element.
This substitution is, however, not uniquely determined by perturbation theory.
One proposal has been made recently by Pisarski.\cite{pisarski} 
We have investigated some similar
$Z_3$ symmetric models. The hope would be to get a description, which is
valid all the way down to the phase transition.

\begin{figure}
\begin{center}
\psfrag{xlabel}[c][c][1][0]{$g^2_3/T$}
\psfrag{ylabel}[c][c][1][0]{$M_S/\sqrt{Tg^2_3}$}
\includegraphics[width=10.3cm]{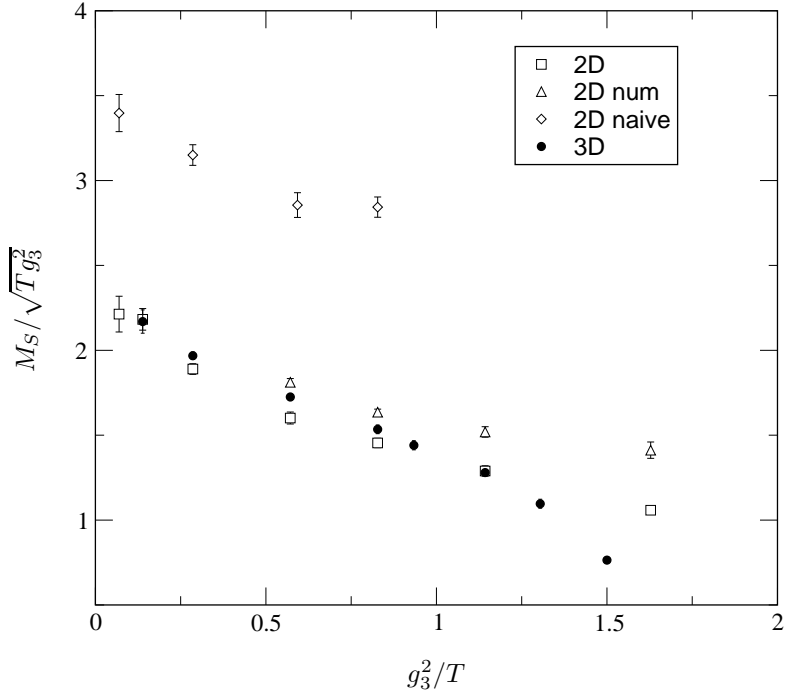}%
\end{center}
\caption{Physical screening masses $M_S$ in units of $g_3\sqrt T$
versus $g_3^2/T$, in $(2+1)D$ (black points) and $2D$ (squares). 
Also shown are the masses obtained 
with the numerical value of the two point coupling
(triangles) instead of its scaling form, and with 
the tree level reduced action(diamonds).}
\end{figure}

To the physical variables $g_3$ and $T$ correspond the lattice parameters
$L_0=1/aT$ and $\beta_3 = 6/ag_3^2$.
It follows that as $a\to 0$, scaling (constant physics)
corresponds to
\bea
L_0\to \infty , \quad \beta _3 \to\infty,\qquad
\tau \equiv \frac {\beta _3}{6L_0}= \frac
{T}{g_3^2}=constant. \label {scaling}
\eea
The dimensionless quantity $\tau$ thus measures temperature in units of
the scale $g_3^2$, high temperature means large $\tau$ values.
From numerical simulations we obtain that $T_c/g_3^2\approx 0.61$ \cite{lego}.
Also, on the lattice, $L_s$ must be kept much larger than the
largest spatial correlation length in lattice units.

The two quantities, which we have studied with precision are the correlation
between Polykaov loops and the spatial string tension. In Fig. 1 is shown 
the screening mass, as defined from the inverse correlation length of
the Polyakov loops. The details of the fit are discussed in the article.
It is important to mention that the best fit is obtained by a simple pole,
although in resummed perturbation theory one might expect a cut, because
two gluons have to be exchanged, as the Polyakov loops are colour singlets.
As can be seen from the figure the agreement between the reduced and the 
full model is very good down to about $1.5T_c$.
Going to the critical temperature, the mass in the full model goes to
zero, while in the reduced model, which does not have the corresponding phase
transition, it stays finite. The data for the full model comes from lattice
calculations with $L_0=4$. We have, however, investigated scale breaking 
effects in the reduced model, and found them to be small.

Another important message is that the terms coming from the quantum effects 
in the reduction are essential for the agreement between the reduced and
the full model. The naively reduced model gives masses, which are more than
50\% larger. 

The spatial Wilson loop in three dimensions is not a static operator, and thus
not predicted to assume the same value at high temperature and in the reduced
two dimensional model. However, there is a finite string tension in both 
models,
and one may expect that the non static corrections to this observable are
small. We have also compared the string tension in the full model to
the string tension in two dimensional $SU(3)$ theory, which can be easily
calculated analytically. In fact one expects
\be
\frac {\sigma _2^0}{g_3^2T}=\frac {2}{3}+\frac {7}{36} \frac
{g_3^2}{T}(aT)^2+\cO \Biggl [(\frac {g_3^2}{T})^2(aT)^4 \Biggr ],
\ee
\begin{figure}[!t]
\begin{center}
\psfrag{ylabel}[c][c][1][0]{$\sqrt{{\sigma}/{Tg_{3}^{2} }}$}
\psfrag{xlabel}[c][c][1][0]{${g_{3}^{2}}/{T}$}
\includegraphics[width=10.3cm]{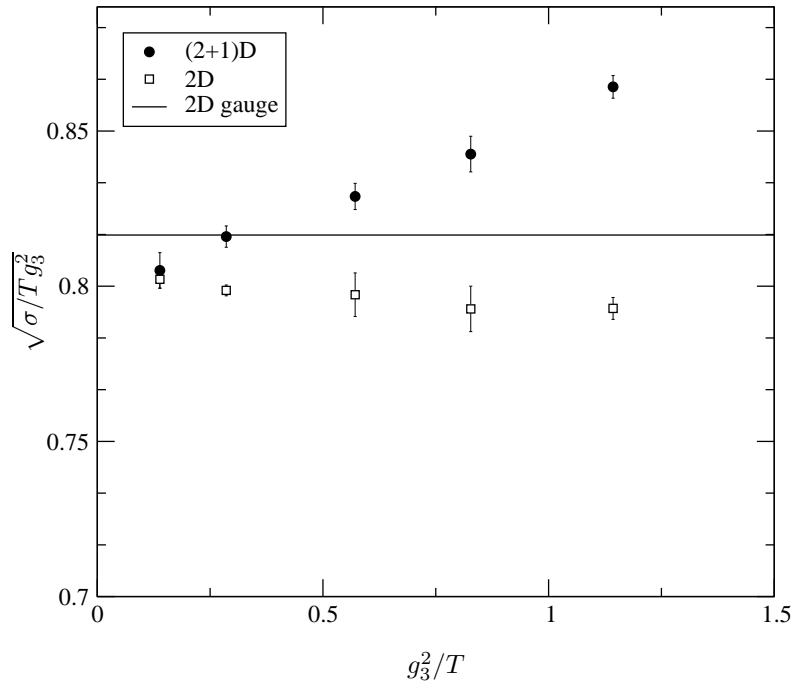}
\end{center}
\caption{The square root of the physical  string tension in units of 
$g_3\sqrt{T}$ as a function of $g_3^2/T$ in (2+1)D (filled circles) 
and 2D (squares). The line denotes the scaling limit $\sqrt {2/3}$
in 2D pure gauge theory. }
\end{figure}
showing that $\sigma _2^0$ scales as $2g_3^2T/3$,
up to scaling violations at finite $T$ of order
$(aT)^2=1/L_0^2$, where $\sigma_2^0$ is the string tension in two dimensional
$SU(3)$. In Fig. 2 is shown the comparison between the spatial string tension
in the full and the reduced model, as well as the analytical result for two
dimensional SU(3). One should note that this is done only for a fixed value
of $L_0 = 4$. In this case we have not estimated scale breaking effects.

\section{Conclusions}

Although the thermodynamics of pure gluon theory is essentially solved
numerically, the same is not true for full QCD. Furthermore, the long
distance properties are even more difficult to extract. Dimensional
reduction gives potentially a simpler and systematic alternative to direct
QCD calculations at high temperature and large distances. We have found
that in $2+1$ dimensional $SU(3)$ it is a very good approximation down
to about $1.5 T_c$, but the dynamics of the phase transition is not
captured by the present approach. It is, however, important to notice that
we can describe both chromoelectric and chromomagnetic quantities consistently,
without extra parameters, which has not been shown in the $3+1$ dimensional 
case.
Systematic extensions of the method would
be very important, in particular since the potential gain in computer
time is enormous.

\section*{Acknowledgments}
I want to thank my collaborators P. Bialas, C. Legeland, A. Morel, 
K. Petrov and T. Reisz
for a very stimulating collaboration, which led to the results described here.
I also thank Xiang-Qian Luo and Eric Gregory for a very interesting conference.

\end{document}